\documentclass[conference]{IEEEtran}
\IEEEoverridecommandlockouts
\usepackage{cite}
\usepackage{amsmath,amssymb,amsfonts}
\usepackage{algorithmic}
\usepackage{graphicx}
\usepackage{textcomp}
\usepackage{xcolor}
\def\BibTeX{{\rm B\kern-.05em{\sc i\kern-.025em b}\kern-.08em
    T\kern-.1667em\lower.7ex\hbox{E}\kern-.125emX}}
    
\IEEEpubid{\makebox[\columnwidth]{979-8-3315-7391-1/25/\$31.00~\copyright~2025 \hfill}
\hspace{\columnsep}\makebox[\columnwidth]{ }}

\begin{document}
\renewcommand\IEEEkeywordsname{Keywords}

\title{Deep Learning–Based Quantum Transport Simulations in Two-Dimensional Materials\\
\thanks{SKP acknowledges NFSG grant from BITS-Pilani, Dubai campus. QG acknowledges the funding support from National Natural Science Foundation of China (No. 12504285) }
}

\author{
\IEEEauthorblockN{Jijie Zou\IEEEauthorrefmark{1}\IEEEauthorrefmark{2}}
\IEEEauthorblockA{
\IEEEauthorrefmark{1}\textit{Centre for Nanoscale Science and Technology,}\\
\textit{Academy for Advanced Interdisciplinary Studies} \\
\textit{Peking University}, 
Beijing, China 
}
\IEEEauthorblockA{
\IEEEauthorrefmark{2}\textit{ AI for Science Institute},
Beijing, China
}
\and 

\IEEEauthorblockN{Zhanghao Zhouyin\IEEEauthorrefmark{3}}
\IEEEauthorblockA{\IEEEauthorrefmark{3}\textit{Department of Physics},
\textit{McGill University},
Montreal, Canada }

\and 

\IEEEauthorblockN{Qiangqiang Gu\IEEEauthorrefmark{4}\IEEEauthorrefmark{1}\IEEEauthorrefmark{5}\IEEEauthorrefmark{6}}
\IEEEauthorblockA{\IEEEauthorrefmark{4} \textit{School of Artificial Intelligence and Data Science}, 
\textit{University of Science and Technology of China},
Hefei, China \\
guqq@ustc.edu.cn} 
\IEEEauthorblockA{
\IEEEauthorrefmark{1}\textit{AI for Science Institute},
Beijing, China
}
\IEEEauthorblockA{
\IEEEauthorrefmark{5}\textit{Suzhou Institute for Advanced Research},
\textit{University of Science and Technology of China}, 
Suzhou, China
}
\IEEEauthorblockA{
\IEEEauthorrefmark{6}\textit{Suzhou Big Data \& AI Research and Engineering Center},
Suzhou, China} 
\and 

\IEEEauthorblockN{Shishir Kumar Pandey\IEEEauthorrefmark{7}\IEEEauthorrefmark{8}}
\IEEEauthorblockA{\IEEEauthorrefmark{7}\textit{Department of General Science}, 
\textit{Birla Institute of Technology and Science, Pilani, Dubai Campus},
Dubai, United Arab Emirates \\
shishir.kr.pandey@gmail.com}
\IEEEauthorblockA{\IEEEauthorrefmark{8}\textit{Department of Physics}, 
\textit{Birla Institute of Technology and Science, Pilani, Hyderabad Campus},
Medchal District, Telangana, India}
}

\maketitle
\IEEEpubidadjcol

\begin{abstract}
Two-dimensional (2D) materials exhibit a wide range of electronic properties that make them promising candidates for next-generation nanoelectronic devices. Accurate prediction of their quantum transport behavior is therefore of both fundamental and technological importance. While density functional theory (DFT) combined with the non-equilibrium Green’s function (NEGF) formalism provides reliable insights, its high computational cost limits applications to large-scale or high-throughput studies. 
Here we present DeePTB-NEGF, a framework that combines a deep learning–based tight-binding Hamiltonians derived learned directly from first-principles calculations (DeePTB) with efficient quantum transport simulations implemented in the DPNEGF package. 
To validate the method, we apply it to three prototypical 2D materials: graphene, hexagonal boron nitride ($h$-BN), and  MoS$_2$. The resulting band structures and transmission spectra show excellent agreement with conventional DFT-NEGF results, while achieving orders-of-magnitude improvement in efficiency. 
These results highlight the capability of DeePTB-NEGF to enable accurate and efficient quantum transport simulations, thereby opening avenues for large-scale exploration and device design in 2D materials.
\end{abstract}


\section{Introduction}
Two-dimensional (2D) materials have attracted tremendous attention owing to their exceptional electronic and transport properties~\cite{geim2007rise,mas20112d,novoselov20162d,mir2020recent}. Their diverse electronic characteristics, ranging from gapless semimetals to wide-gap insulators, make them versatile building blocks for next-generation electronic and optoelectronic devices. A fundamental understanding and accurate prediction of their quantum transport properties are therefore of both scientific and technological significance.

First-principles approaches, such as density functional theory (DFT)~\cite{Kohn1965} combined with the non-equilibrium Green’s function (NEGF) formalism~\cite{keldysh1965diagram,datta1997electronic}, have been widely employed to investigate charge transport in nanoscale systems. While highly reliable and successful in explaining a broad range of experimental phenomena, these methods are computationally demanding, which limits their applicability to realistic device dimensions and high-throughput studies~\cite{zou2024deep}. This challenge highlights the urgent need for alternative strategies that preserve first-principles accuracy while offering substantially improved efficiency.

Deep learning has recently emerged as a powerful tool to bridge this gap. By learning Hamiltonians directly from first-principles data, deep learning models can deliver near-DFT accuracy at a fraction of the computational cost~\cite{guDeep2024,zhouyin2025learning}. When combined with the NEGF framework, this approach enables efficient and accurate quantum transport simulations, thereby opening new opportunities for large-scale and high-throughput device exploration~\cite{burkle2021deep,zhang2025physics,zou2024deep}.

In this work, we demonstrate the capability of DeePTB-NEGF~\cite{zou2024deep},  a deep learning–accelerated quantum transport framework, by applying it to three prototypical 2D materials: graphene, $h$-BN and MoS$_2$. These systems span the full spectrum of electronic behavior including semimetal, insulator and semiconductor  to provide a rigorous testbed for assessing the universality and reliability of the method. Our results show excellent agreement with conventional DFT-NEGF calculations\cite{Brandbyge2002} in reproducing both band structures and transmission spectra, underscoring the potential of deep learning to serve as a scalable and accurate tool for quantum transport  studies in 2D materials.

\section{Methodology}
In this section, we present the overall workflow of our methodology, as illustrated in Fig.~\ref{fig:workflow}. Starting from the atomic structures, DeePTB~\cite{guDeep2024,zhouyin2025learning} is used to train and infer the tight-binding (TB) Hamiltonian, which is subsequently combined with our open-source quantum transport package DPNEGF~\cite{zou2024deep} to simulate electronic transport properties such as the transmission spectrum. The following subsections provide a detailed description of each step.

\subsection{Deep Learning tight-binding Hamiltonian Framework}
\begin{figure}[t]
\centering
\includegraphics[width=8.5 cm]{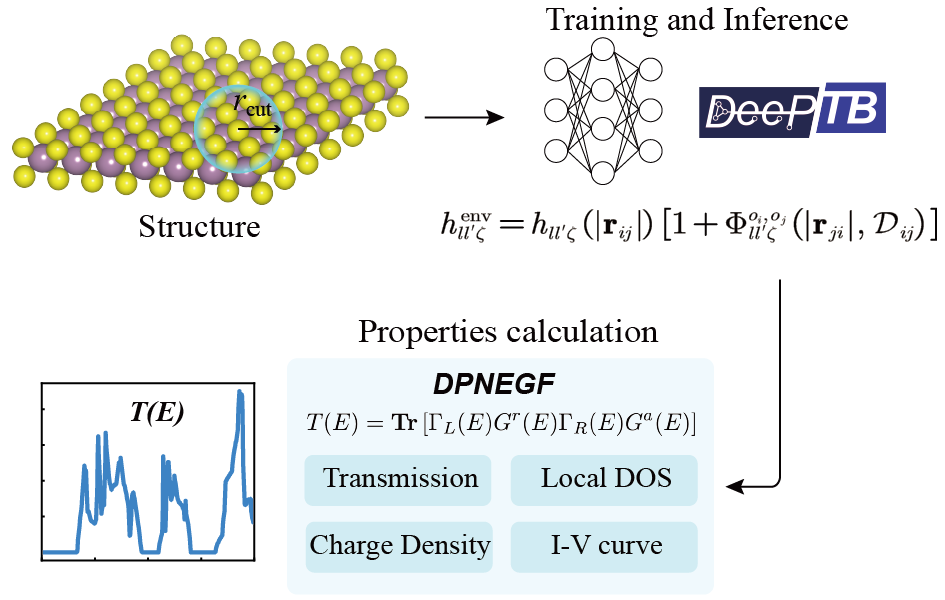}
\caption{Schematic workflow of the DeePTB-NEGF framework applied to two-dimensional materials. Hamiltonian training and inference are performed using DeePTB~\cite{guDeep2024,zhouyin2025learning}, and electronic transport properties are computed with DPNEGF~\cite{zou2024deep}.}
\label{fig:workflow}
\end{figure}

We employ DeePTB to predict TB Hamiltonians for the target systems. DeePTB learns the mapping from local atomic environments to Hamiltonian matrix elements, as illustrated in Fig.~\ref{fig:workflow}. Neural networks generate environment-dependent corrections to Slater–Koster (SK) integrals, improving accuracy beyond conventional two-center TB models while preserving the sparsity of the Hamiltonian. By training in structures with various local atomic environments, DeePTB can reliably predict electronic structures of previously unseen systems. For more details, please refer to Ref.~\cite{guDeep2024,zhouyin2025learning}.

\subsection{Integration with DPNEGF}
For quantum transport simulations, DeePTB are coupled with NEGF formalism to compute transport properties. Within this formalism, the retarded Green’s function at energy $E$ is expressed as,
\begin{equation}
    G^r(E) = \big[E I - H - \Sigma_L^r(E) - \Sigma^r_R(E)\big]^{-1}
\end{equation}
where $H$ denotes the predicted Hamiltonian, and $\Sigma_{L,R}^r(E)$ are the retarded self-energies describing the semi-infinite electrode. The transmission spectrum is then evaluated as,
\begin{equation}
    T(E) = \mathrm{Tr}\big[\Gamma_L(E) G^r(E) \Gamma_R(E) G^a(E)\big]
\end{equation}
with $\Gamma_{L,R}(E) = i[\Sigma_{L,R}(E) - \Sigma_{L,R}^\dagger(E)]$.  
The NEGF calculation has been implemented in DPNEGF~\cite{zou2024deep,dpnegf-git}.

\subsection{Systems Studied}
To demonstrate the reliability and generality of the approach, we focus on three prototypical two-dimensional (2D) materials: graphene, $h$-BN and MoS$_2$. These systems not only represent the most widely studied 2D systems but also collectively span the full range of electronic behaviors—semimetal, semiconductor, and insulator. Their diversity therefore provides an ideal and stringent testbed for the framework in 2D material systems.
The DFT reference data for graphene and $h$-BN were generated using the SIESTA package~\cite{Soler2002}, while those for MoS$_2$ were generated using the ABACUS package~\cite{chen2010systematically,li2016large,lin2024ab}. Exchange–correlation effects were described by the Perdew–Burke–Ernzerhof (PBE) functional within the generalized gradient approximation (GGA). 
A vacuum spacing of more than 15~\AA~was used to eliminate spurious interlayer interactions.  
The DFT band structures are extracted from these calculations, which serve as the training labels for DeePTB model. Transmission spectra were then computed using DPNEGF, and compared with the results from the established DFT-NEGF approach~\cite{Brandbyge2002,papior2017improvements}, enabling a one-to-one comparison of the two approaches. This procedure provides a systematic validation of the framework across 2D materials with distinct electronic structures.

\section{Results}
In this section, we firstly benchmark the calculated transmission spectra of graphene and $h$-BN against conventional DFT-NEGF implementation (TranSIESTA~\cite{Brandbyge2002,papior2017improvements}),  confirming  the accuracy and efficiency of our approach. We then demonstrate the unique flexibility of DeePTB-NEGF by studying MoS$_2$, where the DeePTB Hamiltonian is constructed from learning the electronic structure computed with the ABACUS package~\cite{chen2010systematically,li2016large,lin2024ab}, a DFT code without a native transport module. This result showcases the power of our framework to decouple electronic structure calculations from transport simulations, effectively enabling quantum transport studies for a broad range of DFT packages that lack integrated NEGF functionalities.

For all three systems, the transmission spectra are computed over an energy window from $-10$~eV to $10$~eV with a step size of $0.05$~eV, using 100 $k$-points sampled in the in-plane Brillouin zone perpendicular to the transport direction.

\begin{figure}[t]
\centering
\includegraphics[width=7.2 cm]{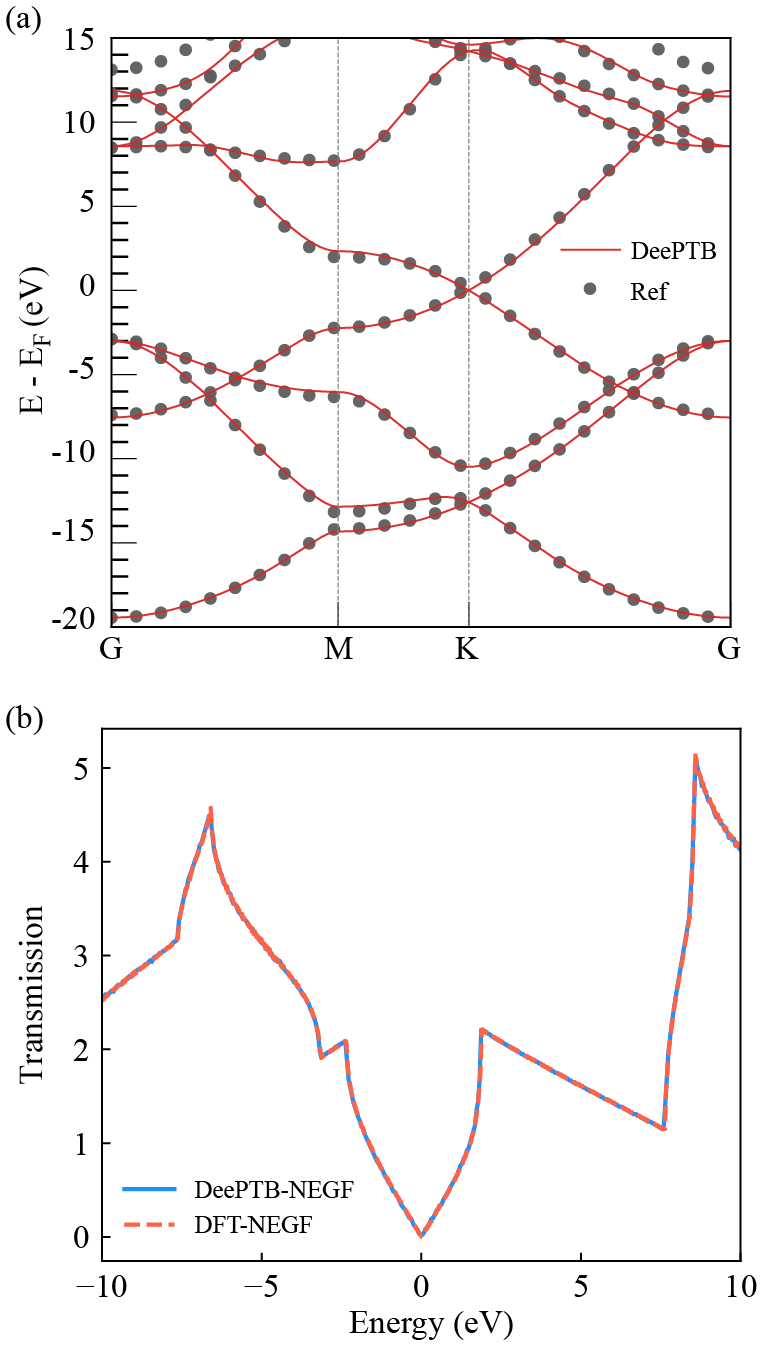}
\caption{DeePTB-NEGF simulation results for graphene. (a) Comparison of band structures obtained from DeePTB and DFT (SIESTA). (b) Comparison of transmission spectrum from DeePTB-NEGF and DFT-NEGF.}
\label{fig:Gr}
\end{figure} 
\subsection{Graphene}
Graphene serves as a prototypical test case due to its unique semimetallic nature and linear dispersion near the Dirac point. Fig.~\ref{fig:Gr}(a) compares the band structures obtained from DeePTB and DFT package SIESTA~\cite{Soler2002}. The two sets of results are in excellent agreement, accurately reproducing the characteristic gapless Dirac cone at the Fermi level as well as the overall dispersion across the Brillouin zone. This demonstrates the ability of DeePTB to capture both valence bands and low-energy conduction bands with high fidelity. 

The two transmission spectra (see Fig.~\ref{fig:Gr}(b)) exhibit an almost complete overlap across the considered energy window, demonstrating quantitative agreement in the fine features near the Dirac point. Notably, DeePTB-NEGF delivers this level of accuracy with substantially reduced computational cost, underscoring its promise as a scalable tool for transport simulations in metallic and semimetallic 2D materials.

\begin{figure}[htbp!]
\centering
\includegraphics[width=7.2 cm]{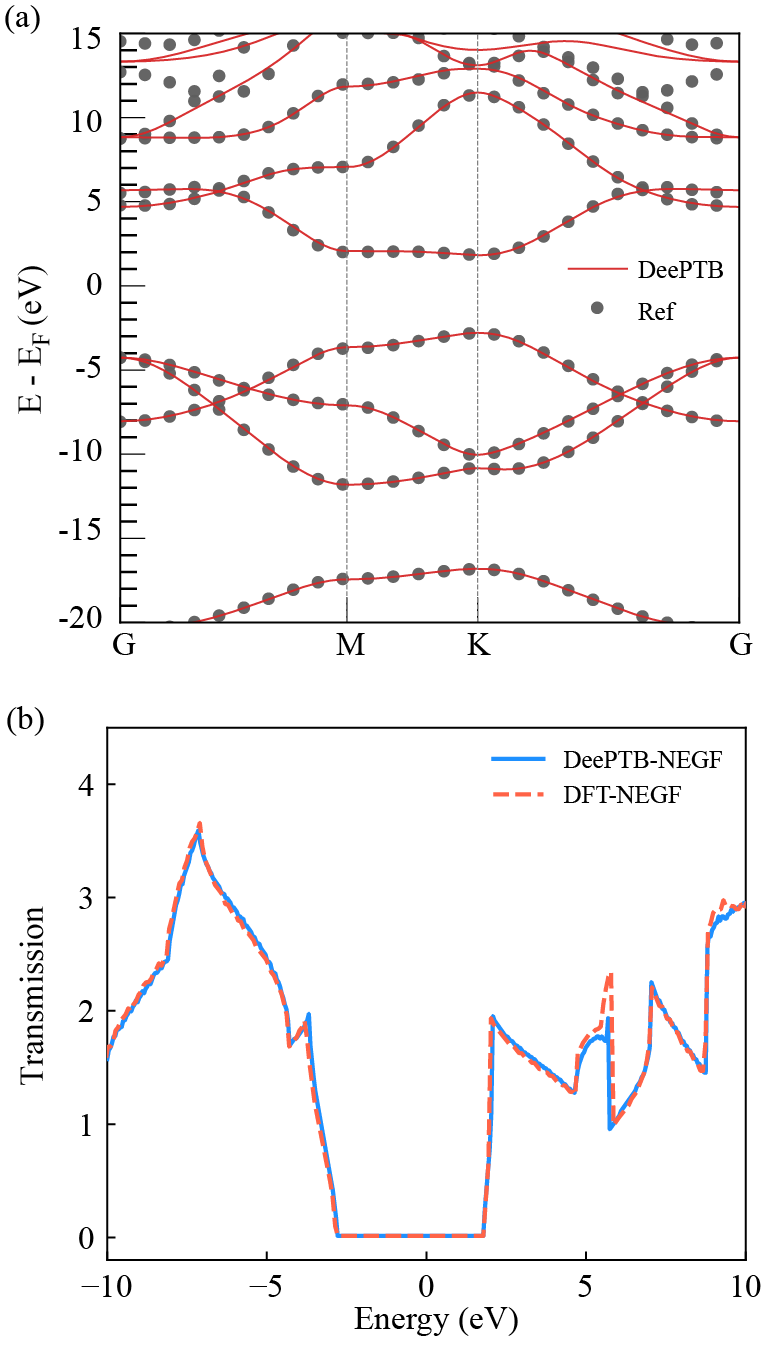}
\caption{DeePTB-NEGF simulation results for $h$-BN. (a) Comparison of band structures obtained from DeePTB and DFT (SIESTA). (b) Comparison of transmission spectrum from DeePTB-NEGF and DFT-NEGF.}
\label{fig:hBN}
\end{figure} 

\subsection{Hexagonal boron nitride}
Hexagonal boron nitride ($h$-BN) provides a complementary benchmark to graphene due to its wide-bandgap insulating character. Fig.~\ref{fig:hBN}(a) shows the band structures from DeePTB compared with those obtained from DFT (SIESTA). The two are in close agreement, correctly reproducing the large band gap and the overall dispersion of both the valence and conduction bands. This confirms the ability of DeePTB to handle wide-gap insulators with accuracy comparable to first-principles calculations.

The transport properties of $h$-BN, represented by the transmission spectrum in Fig.~\ref{fig:hBN}b, further validate this consistency. DeePTB-NEGF reproduces the DFT-NEGF results obtained from TranSIESTA~\cite{Brandbyge2002,papior2017improvements} correctly capturing the vanishing transmission within the band gap and the onset of conduction channels at higher or lower energies. Such agreement indicates that the deep learning–based approach not only describes metallic and semimetallic systems, as in graphene, but also extends reliably to insulating 2D materials.


\subsection{Molybdenum Disulfide}
As a representative semiconductor, monolayer MoS\textsubscript{2} offers an intermediate case between graphene and $h$-BN. Unlike graphene (semimetal) and $h$-BN (insulator), MoS\textsubscript{2} possesses a sizable direct band gap, making it a prototypical 2D material for transistor applications~\cite{ganatra2014few,lembke2015single}.

\begin{figure}[htbp!]
\centering
\includegraphics[width=7.2 cm]{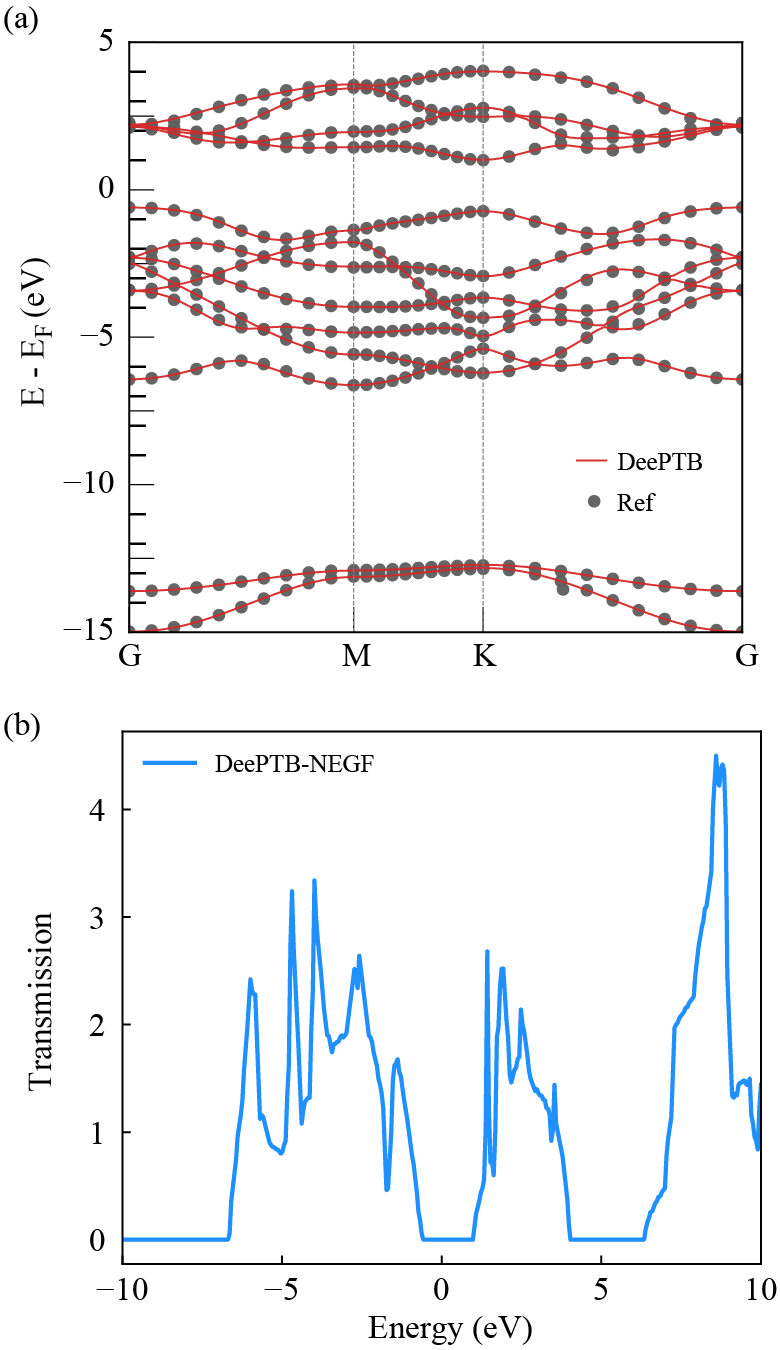}
\caption{DeePTB-NEGF simulation results for MoS\textsubscript{2}. (a) Comparison of band structures obtained from DeePTB and DFT (ABACUS). (b) The transmission spectrum from DeePTB-NEGF.}
\label{fig:MoS2}
\end{figure}

Fig.~\ref{fig:MoS2}(a) compares the band structure obtained from DeePTB with first-principles calculations performed using the ABACUS package. The DeePTB model faithfully reproduces the conduction and valence band dispersions near the K and $\Gamma$ points, as well as the magnitude of the band gap. This agreement demonstrates that the deep learning framework can capture the essential features of semiconducting 2D materials, even when trained on data from different first-principles codes.

The transmission spectrum computed with DeePTB-NEGF is shown in Fig.~\ref{fig:MoS2}(b). In the absence of a DFT-NEGF reference—since ABACUS does not currently provide NEGF capabilities—the spectrum is derived from the DeePTB-fitted Hamiltonian. The results correctly reflect the semiconducting nature of MoS\textsubscript{2}, with negligible transmission within the gap and the gradual emergence of conducting channels at higher energies. Together with the graphene and $h$-BN cases, the MoS\textsubscript{2} results highlight the generality of DeePTB-NEGF across the spectrum of electronic behaviors in 2D materials.

\subsection{Computational Efficiency Benchmark}

In addition to accuracy, computational efficiency is a key advantage of the DeePTB-NEGF framework. To illustrate this, we benchmark the simulation time for graphene and $h$-BN by computing the $\Gamma$-point transmission. Both the conventional DFT-NEGF approach and DeePTB-NEGF are employed on the same computational resources (32-core CPU node) to calculate the transmission from $-2.5$~eV to $2.5$~eV with a step of 0.025~eV. Both comparisons are performed on a system containing 48 atoms.

\begin{figure}[htbp!]
\centering
\includegraphics[width=8 cm]{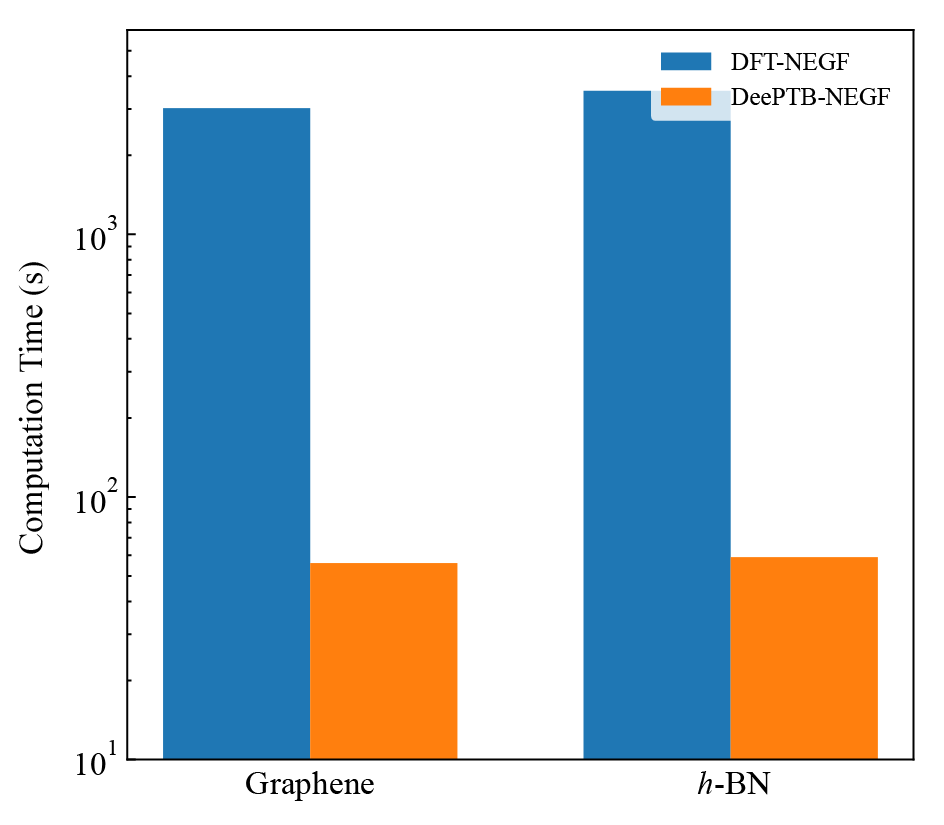}
\caption{Efficiency comparison of DeePTB-NEGF and DFT-NEGF for graphene and $h$-BN by computing the $\Gamma$-point transmission spectrum. DeePTB-NEGF achieves over an order-of-magnitude speed-up in both cases.}
\label{fig:time_bench}
\end{figure}

Fig.~\ref{fig:time_bench} summarizes the wall-clock time required for Hamiltonian construction and transmission calculations. While DFT-NEGF requires 3024~seconds for graphene and 3521~seconds for $h$-BN to compute the full transmission spectrum, DeePTB-NEGF completes the same task in 56~seconds and 59~seconds, respectively, achieving more than an order-of-magnitude speed-up. This efficiency gain stems from the fact that Hamiltonian generation via deep learning bypasses the need for self-consistent calculations for each configuration.

Although the preparation of training data and the training process require additional time, DeePTB-NEGF can infer Hamiltonians for large, previously unseen structures, dramatically reducing the computational cost for high-throughput materials exploration and rapid prototyping of 2D electronic devices~\cite{guDeep2024,zhouyin2025learning,zou2024deep}. While graphene and $h$-BN are used here as  representative examples, similar efficiency improvements are expected for other 2D materials. 
For larger systems, the efficiency improvement becomes even more pronounced, as the computational cost of self-consistent calculations increases drastically with system size, as demonstrated in Ref.~\cite{zou2024deep}.

\section{Conclusion}
In this work, we have demonstrated the capability of the DeePTB-NEGF framework to perform accurate and efficient quantum transport simulations in 2D materials. Benchmark results for graphene, $h$-BN, and MoS\textsubscript{2} show excellent agreement with conventional DFT-NEGF calculations, while achieving orders-of-magnitude speed-up. By combining deep learning-predicted TB Hamiltonians with the NEGF formalism, DeePTB-NEGF enables scalable simulations for large and previously unseen structures, providing a practical route for high-throughput exploration and rapid prototyping of 2D electronic devices. These results highlight the potential of deep learning-based quantum transport methods to accelerate the design and discovery of next-generation nanoscale materials and devices.

\bibliographystyle{IEEEtran}
\bibliography{ref}

\end{document}